\documentclass[10pt,a4paper,twocolumn]{article}
\usepackage[utf8]{inputenc}
\usepackage{amsmath}
\usepackage{amsfonts}
\usepackage{amssymb}
\usepackage{graphicx}
\usepackage{url}
\usepackage{authblk}
\usepackage[margin=2.5cm]{geometry}

\title{Human Mobility and Predictability enriched by Social Phenomena Information}

\author[1,2]{Nicolas B. Ponieman}
\author[2]{Alejo Salles}
\author[1]{Carlos Sarraute}
\affil[1]{Grandata Labs, Argentina}
\affil[ ]{ \texttt{ \{nico,charles\}@grandata.com } }
\affil[2]{Physics Dept., Universidad de Buenos Aires, Argentina}
\affil[ ]{ \texttt{ alejo@df.uba.ar } }
\date{}

\begin{document}
\maketitle
\thispagestyle{empty}
\pagestyle{empty}

\section{Introduction}

The information collected by mobile phone operators can be considered
as the most detailed information on human mobility across a large part
of the population \cite{song2010limits}.
The study of the dynamics of human mobility using the collected
geolocations of users, and applying it to predict future users' locations,
has been an active field of research in recent 
years \cite{domenico2012interdependence,lu2012predictability}.

In this work, we study the extent to which social phenomena are reflected in mobile phone data, 
focusing in particular in the cases of urban commute and major sports events. 
We illustrate how these events are reflected in the data, and show how information 
about the events can be used to improve predictability in a simple model for a mobile 
phone user's location.

\section{Mobile Data Source}

Our data source is anonymized traffic information from a mobile operator in Argentina, focusing mostly in the Buenos Aires metropolitan area, over a period of 5 months. 
We use Call Detail Records (CDR) including time of the call, users involved, direction of the call (incoming/outgoing), the antenna used in the communication, and its position.
The raw data logs contain around 50 million calls per day.
CDRs are an attractive source of location information since they are collected 
for all active cellular users (about 40 million users in Argentina),
and creating additional uses of CDR data incur little marginal cost.

\section{Mobility Model}

\begin{figure}
	\centering
	\includegraphics[width=0.5\textwidth]{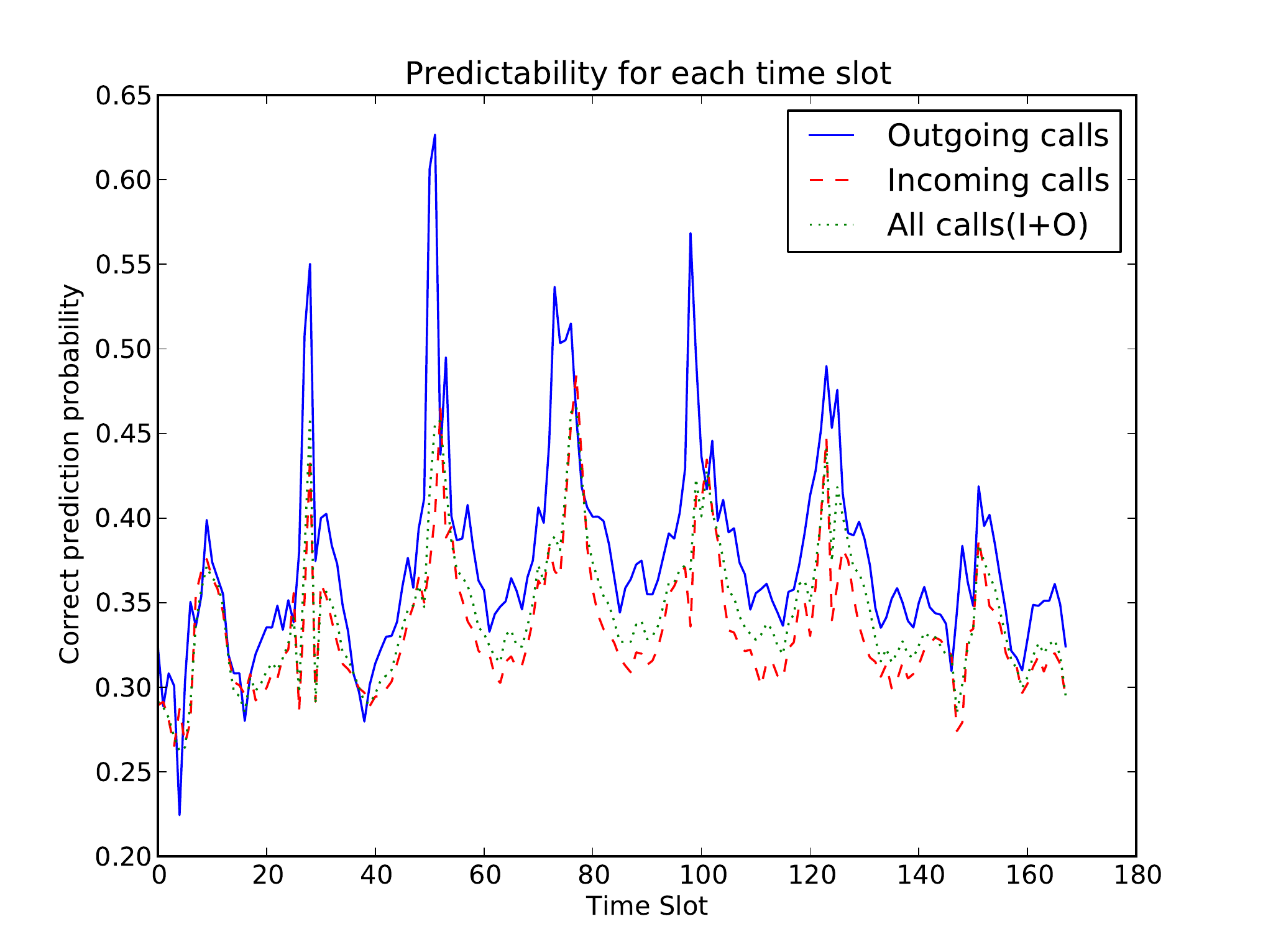}
	\caption{Users' location predictability by time slot. 
	Blue: Outgoing calls. Red: Incoming calls. Green: All calls.}
	\label{fig:predictability}
\end{figure}

To predict a user's position, we use a simple model based on previous most frequent locations. 
In order to compute these locations, we split the week in time slots, one for each hour, totalizing $7*24 = 168$ slots per week. 
Since humans tend to have very predictable mobility 
patterns \cite{song2010limits,gonzalez2008understanding,jiang2012clustering}, 
this simple model turns out to give a good predictability baseline, achieving an average of around $35 \%$ correct predictions for a period of 2 weeks, training with 15 weeks of data, including peaks of above $50 \%$ predictability. 
This model was used as a baseline in~\cite{cho2011friendship}, with which our results agree. 
In Figure~\ref{fig:predictability} we show the average predictability for all time slots.
 
It is important to notice that it is possible to improve the accuracy of this simple model 
by clustering antennas, instead of defining each antenna as a location.

\begin{figure*}[th]
\begin{minipage}{.33\textwidth}
  \centering
  \includegraphics[width=0.95\textwidth]{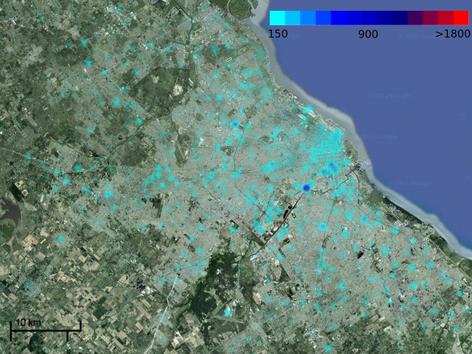}
  (a) 6 a.m.
\end{minipage}
%\hfill
\begin{minipage}{.33\textwidth}
  \centering
  \includegraphics[width=0.95\textwidth]{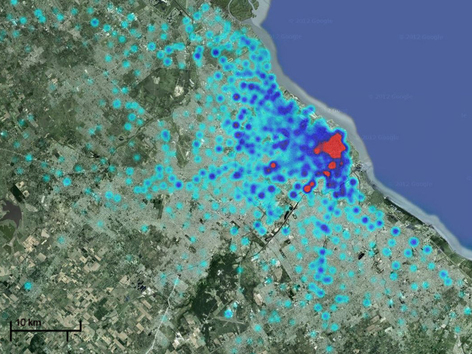}
  (b) 8 a.m.
\end{minipage}
\begin{minipage}{.33\textwidth}
  \centering
  \includegraphics[width=0.95\textwidth]{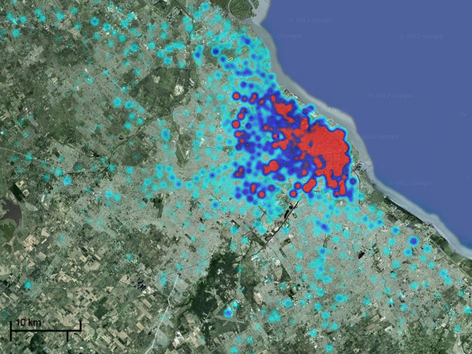}
  (c) 10 a.m.
\end{minipage}

\begin{minipage}{.33\textwidth}
  \centering
  \includegraphics[width=0.95\textwidth]{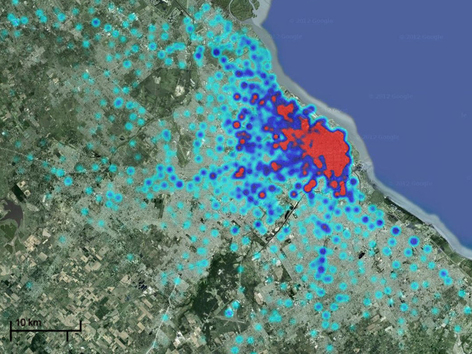}
  (d) 5 p.m.
\end{minipage}
%\hfill
\begin{minipage}{.33\textwidth}
  \centering
  \includegraphics[width=0.95\textwidth]{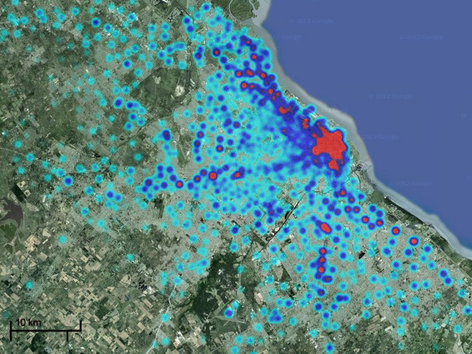}
  (e) 7 p.m.
\end{minipage}
\begin{minipage}{.33\textwidth}
  \centering
  \includegraphics[width=0.95\textwidth]{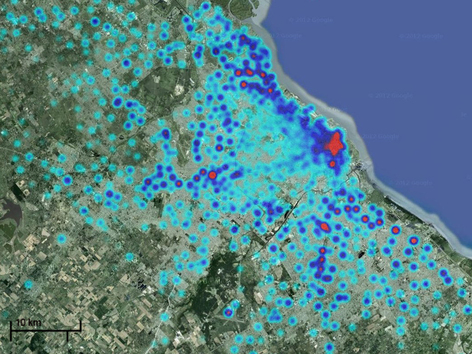}

  (f) 8 p.m.
\end{minipage}

  \caption{\label{fig:commute} Commute to Buenos Aires city from the surrounding areas on a weekday,
   for different hours.
    Red color corresponds to a higher number of calls, whereas blue corresponds to an intermediate number of calls and light blue to a smaller one.}
\vspace{-0.0cm}
\end{figure*}

\section{Urban Commute}

The phenomenon of commuting is prevalent in large metropolitan areas (often provoking upsetting traffic jams and incidents), and naturally appears in mobile phone data.
For instance, in \cite{isaacman2011identifying} the authors study commute distances
in Los Angeles and New York areas.
Mobile data can lead to quantification of this phenomenon in terms of useful quantities, which are much harder to measure directly. 
We include a series of call patterns illustrating the Buenos Aires commute in 
Figure~\ref{fig:commute}.
%\footnote{\label{note1} Greater resolution versions
%of these maps, as well as additional figures, are available in 
%the Labs section at \url{www.grandata.com}.}. 

From the data, we can estimate the radius of the commute (the average distance traveled by commuters). Considering the two most frequently used antennas as the important 
places for each user (home and work, see \cite{csaji2012exploring}), we get an
average commute radius of $7.8$ km (as a comparison, the diameter of the city
is about 14 km, and the diameter of the considered metropolitan area is 90 km).

\section{Sports Events}

\begin{figure*}[th]
\begin{minipage}{.245\textwidth}
  \centering
  \includegraphics[width=0.95\textwidth]{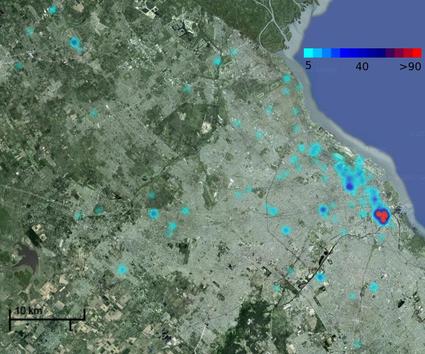}
  (a) 5 hours before
\end{minipage}
%\hfill
\begin{minipage}{.245\textwidth}
  \centering
  \includegraphics[width=0.95\textwidth]{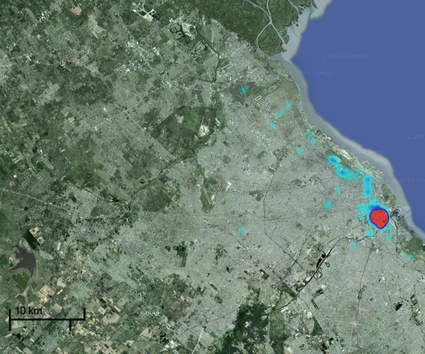}
  (b) 1 hour before
\end{minipage}
\begin{minipage}{.245\textwidth}
  \centering
  \includegraphics[width=0.95\textwidth]{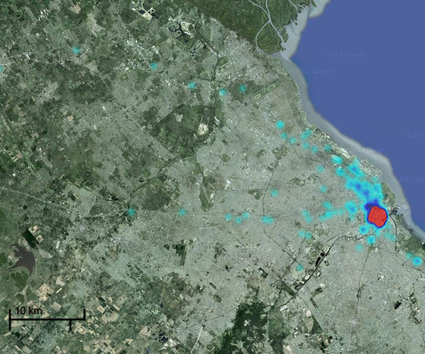}
  (c) 1 hour after
\end{minipage}
%\hfill
\begin{minipage}{.245\textwidth}
  \centering
  \includegraphics[width=0.95\textwidth]{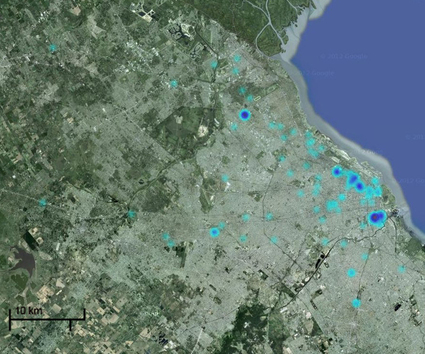}
  (d) 3 hours after
\end{minipage}
  \caption{\label{fig:boca_juega} Convergence to Boca Juniors stadium on hours prior to a soccer match, and dispersal after its end. 
        Red color corresponds to a higher number of calls, whereas blue corresponds to an intermediate number of calls and light blue to a smaller one.}
\vspace{-0.0cm}
\end{figure*}

As in the urban commute case, we study human mobility in sports events as seen through mobile phone data. In Figure~\ref{fig:boca_juega}, we show how assistants to a Boca Juniors soccer match converge to the stadium in the hours prior to the game, and disperse outwards.
% \footnotemark[\ref{note1}].

Note that postselecting the users attending the event necessarily produces the effect of having no calls outside the chosen area during the match, however, the convergence pattern observed is markedly different from the one seen for the same time slot of the week on a day with no match.

\subsection*{Improving Predictability with External Data}

So far, our results allow us to understand (and quantify) social events through the analysis of mobile phone data. This understanding can be in turn used to improve the mobility model. Social relations among individuals have been used to improve predictability in mobility models before, as in~\cite{cho2011friendship}, where social links learned from the mobile phone records are used to this end. Here, instead of peer to peer links learned from the mobile data, we show how an external data source can be used to improve the model.

We illustrate this effect using as proof of concept the case study of soccer matches.
By taking the soccer fixture, we tag users as ``Boca Juniors fans'' if they make calls using antennas around the stadium and time slot where Boca plays for three consecutive matches (which include both home and away matches). 
Using this tagging, we can dramatically improve predictability for this group of Boca fans, even predicting positions that had never been visited by a user before. The predictability of the model for these users considering the fixture data rises for the matches to $38\%$ -- which doubles the $19\%$ accuracy achieved by our previous model for the same set.
Moreover, the initial model is only able to make predictions in $63\%$ of events in the given set (as a consequence of a lack of information from the training set data), whereas the socially enriched model tries to predict $100\%$ of the events during match days, which make the previous results even more significant.

In order to understand these results, we illustrate with an example where the enriched model outperforms the simple model:
	the simple model would rarely predict a user's location on a different city, whereas the enriched model would do so if the user is a Boca fan, and Boca has an away match in that city.

\section{Conclusion}

We illustrated how social phenomena can be studied through the lens of mobile phone data, which can be used to quantify different aspects of these phenomena with great practicity. Furthermore, we showed how including external information about these phenomena can improve the predictability of human mobility models.

Although we showed this in a specific case as a proof of concept experiment, we note that this procedure can be extended to other settings, not restricted to sports but including cultural events, vacation patterns and so on (see \cite{lu2012predictability} for a specially relevant application). The tagging obtained is useful on its own and is of great value for mobile phone operators.  
The big challenge in this line of work is to manage to include external data sources
in a systematic way.

Lastly, the tag-based predictions can be taken to the community level. Defining, for instance, the ``Boca Juniors fans'' community, we can predict that if some users of this community make or receive calls in a certain location, other users in the community will do it as well.

\bibliographystyle{unsrt}

\bibliography{mobility}

\end{document}